%% file: kratzer_furrer_pittavino.tex
\documentclass{svproc}
\usepackage{url}
\usepackage{graphicx}        
\usepackage{multicol}        

\usepackage{subcaption}
\usepackage{cite}

\usepackage[bottom]{footmisc}

\usepackage{bigints}

\usepackage{bm}

\newcommand{\btheta}{{\bm{\theta}}}

\newcommand{\X}{{\textbf{\textit{X}}}}

\newcommand{\Abn}{{\mathcal{A}}}
\newcommand{\Bay}{{\mathcal{B}}}

\newcommand{\Str}{{\mathcal{S}}}
\newcommand{\Par}{{\mathbf{Pa}_j}}

\begin{document}
\mainmatter              
\title{Comparison between Suitable Priors for Additive Bayesian Networks}
\titlerunning{Suitable priors for ABN}  
%
\author{Gilles Kratzer, Reinhard Furrer and Marta Pittavino}
\authorrunning{Gilles Kratzer et al.} 
%
\tocauthor{Gilles Kratzer, Reinhard Furrer and Marta Pittavino}
\institute{Department of Mathematics, University of Zurich, Zurich, Switzerland,\\
 \email{gilles.kratzer@math.uzh.ch},\\
\and 
Department of Mathematics and Department of Computational Science, University of Zurich, Zurich, Switzerland,\\
\email{reinhard.furrer@math.uzh.ch},\\
\and 
Research Center for Statistics (RCS), Geneva School of Economy and Management (GSEM), University of Geneva, Geneva, Switzerland,\\
 \email{marta.pittavino@unige.ch}
}

\maketitle              

\begin{abstract}
Additive Bayesian networks  are types of graphical models that extend
 the usual Bayesian generalized linear model to multiple dependent variables through the
 factorisation of the joint probability distribution of the underlying variables.
 When fitting an ABN model, the choice of the prior of the parameters is of crucial importance.
 If an inadequate prior -- like a too weakly informative one -- is used, data separation and data sparsity lead to issues in the model selection process. In this work a simulation study between two weakly and a strongly informative priors is presented.
 As weakly informative prior we use a zero mean Gaussian prior with a large
 variance, currently implemented in the R-package \emph{abn}. The second prior belongs to the Student's \emph{t}-distribution, specifically designed for logistic regressions and, finally,  the strongly informative prior is again Gaussian with mean equal to true parameter value and a small variance. 
 We compare the impact of these priors on the accuracy of the learned additive Bayesian network in function of different parameters.
 We create a simulation study to illustrate Lindley's paradox based on the prior choice.
 We then conclude by highlighting the good performance of the informative Student's \emph{t}-prior and the limited impact of the Lindley's paradox.
 Finally, suggestions for further developments are provided.
 
\end{abstract}

\section{Introduction to ABN}
\label{sec:1}
Additive Bayesian network (ABN) models are a special type of   Bayesian network (BN) models, where
each node in the graph comprises a generalized linear model (GLM).
All types of BN models consist of two reciprocally dependent parts: a
qualitative one (the structure) and a quantitative one (the model parameters).
BN models are statistical models that derive a directed acyclic graph (DAG) from empirical data,
describing the dependency structure of the random variables. The DAG is the graphical representation of the joint
probability distribution of all random variables represented by the data. The model parameters stem from  the local probability distribution of all the variables in the network.  

In the last few decades, BN modelling has been widely used in biomedical science and in systems biology to analyse multi-dimensional data \cite{Jansen2003,Dojer2006,Poon2007,Djebbari2008,Hodges2010}.
However, it is only in the last few years that ABN models have been applied to the veterinary epidemiology field as a result of their ability to
generalize standard regression methodologies \cite{Lewis2012, Hartnack2016, Pittavino2017a}.

Relevant technical details of ABN models are presented in Section~\ref{sec:2}. 
Section~\ref{sec:3} explains the issue of data separation and Lindley's paradox and highlights the importance of appropriate prior choice. 
Section~\ref{sec:4} reports the results of a simulation study underpinning the necessity of careful prior selection with respect to data separation and Lindley's paradox. 
We conclude the article in Section~\ref{sec:5} with future research directions.

\section{Theory of Additive Bayesian Networks  in a Nutshell}
\label{sec:2}
A BN for a set of random variables $\X = \{X_1,\dots, X_n\}$ consists of:
\begin{itemize}
\item A DAG structure $\Str = (\mathit{V},\mathit{E})$, where $\mathit{V}$ is a finite set of
nodes and $\mathit{E}$ is a finite set of directed edges between the nodes. A DAG is \emph{acyclic}; hence, the edges in $\mathit{E}$ do not form
directed cycles.
A random variable $X_j$ corresponds to each node $j \in \mathit{V} = \{1, \dots, n\}$ in the graph. We do not distinguish between a variable $X_j$ and
the corresponding node $j$.
\item A node $k$ is said to be a \emph{parent} of a node $j$ if the edge set $E$ contains an edge from $k$ to $j$.
 A set of parents for a node $j$ is denoted by $\Par$. 
$P_j$ indicates the total number of parents for a node $j: \,\, \dim(\Par)=P_j\geq0$.
$P_j=0$ for orphan nodes.
\item A set of local probability distributions for all
variables in the network, denoted~$\btheta_\Bay$. Each node $j$, with parent set $\Par$, is parametrized by a local probability
distribution: $P(X_j\mid \Par)$.
\end{itemize}
Edges represent both \textit{marginal} and \textit{conditional dependencies}.
The main role of the network structure is to express the conditional
independence relationships among the variables in the model
through graphical separation, thus specifying the factorization of
the global probability distribution:
 $$\displaystyle P(\X)=\prod_{j=1}^{n}P(X_j\mid \Par).$$

We denote a BN model $\Bay$ for a set of random variables $\X$ by a pair $\Bay=(\Str, \btheta_\Bay)$.
The DAG $\Str$ defines the \emph{structure}, and $\btheta_\Bay$ the \emph{parametrization} of the model.
In order to specify a model $\Bay$ for $\X$, we must therefore specify a DAG structure and a set of local probability distributions.

\begin{figure}
\begin{flushright}
\includegraphics[scale=.53]{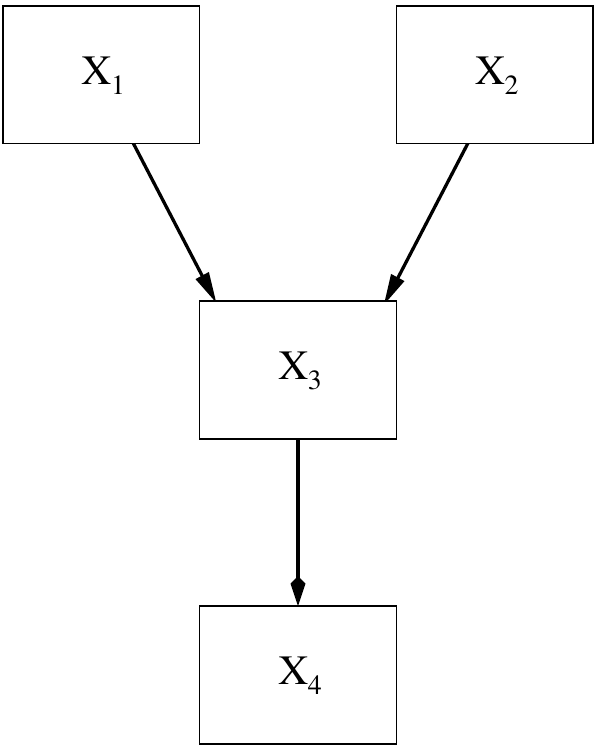}\hspace*{0.3cm}
\caption{A binary additive Bayesian network model $\Abn$ for four random variables ($n=4$, $S_j=2$, $j=1,\dots,n$).}
\label{BN_Col}
\end{flushright}
\hspace*{0.3cm}
\raisebox{3cm}[0pt][0pt]{
\parbox[c]{11.5cm}{
$X_1$ is (conditionally) independent of $X_2$:\\ 
$P(X_1=1)=\theta_1$, $\log\Bigl(\displaystyle\frac{\theta_{1}}{1-\theta_{1}}\Bigr)=\beta_{1,0}$\\
$X_2$ is (conditionally) independent of $X_1$:\\
$P(X_2=1)=\theta_2$, $\log\Bigl(\displaystyle\frac{\theta_{2}}{1-\theta_{2}}\Bigr)=\beta_{2,0}$\\
$X_3$ is conditionally dependent jointly upon $X_1$ and $X_2$:\\
$P(X_3=1\mid X_2,X_3)=\theta_3$, $\log\Bigl(\displaystyle\frac{\theta_{3}}{1-\theta_{3}}\Bigr)=\beta_{3,0}+\beta_{3,1}X_1+\beta_{3,2}X_2$\\
$X_4$ is conditionally dependent upon $X_3$:\\
$P(X_4=1\mid X_3)=\theta_4$, $\log\Bigl(\displaystyle\frac{\theta_{4}}{1-\theta_{4}}\Bigr)=\beta_{4,0}+\beta_{4,1}X_3$
\vspace{1.3cm}
}}
\end{figure}

In the purely discrete case, an additive BN $\Abn$ consists of a BN $\Bay$ that generalizes the multinomial logistic regression model. 
Let $S_j$ be the number of states of the variable $X_j$, and $s \in \{1,\dots, S_j\}$ the corresponding set of indexes.
Let $C_j = \prod_{p:\,X_p \in \Par} S_p$ be the number of configurations of $\Par$ and $c \in \{1, \dots, C_j\}$ indicates
the corresponding set of indexes for the different parents configurations of $\Par$.
Let $X_j=s$ indicate a possible observation for $X_j$. Hence, let $P(X_j = s\mid  \Par = c)$
be the probability that $X_j = s$, given the $c$-th parent configuration of $\Par$, denoted by the multinomial parameter $\theta_{jcs}$.
The multinomial logistic regression model 
can be integrated into a BN $\Bay$ by modelling each of its conditional probability table
$P(X_j=s\mid  \Par = c)=\theta_{jcs}$ with a multinomial logistic regression model, where $X_j$ is progressively the outcome variable and the resulting regression design matrix is constructed from $\Par$, as showed in \cite{Rijmen2008} and in detail in Figure \ref{BN_Col}. 


\vspace*{-0.5cm}

\section{Data separation and Lindley's paradox}
\label{sec:3}
The data separation  arises when a linear combination of predictors predicts perfectly the outcome and is surprisingly common in applied logistic regression. Data separation induces estimation problems for the entire model, not only for the parameters directly involved.

Due to the large number of models necessary to evaluate, data separation is a serious concern when 
modelling  discrete data with an ABN. The separation occurs when the data set is too small to observe events with low probabilities. Therefore, the smaller the sample size the higher is the probability of not observing given instances which have a low probability. The issue is intensified  with increasing complexity of the model. A popular solution is to remove predictors until the design matrix becomes fully ranked. However, this often leads to the deletion of the strongest predictors which is not desirable, especially in the context of ABN \cite{zorn2005solution}. 
Alternatively, the natural ``Bayesian'' solution is to use a prior, which will drive the posterior whenever data separation arises. 
Multiple prior distributions have been proposed to tackle this issue. A notably one is the Jeffreys prior  \cite{firth1993bias} which is, however,  hard to interpret as a prior information. Indeed, Jeffreys prior is not parametrised on the scale of the parameter. Moreover,  when applied to sparse data the prior may lead to poor numerical results. 
As a result, dedicated priors have been developed which are weakly informative enough to be used in a general context and which can still drive the posterior if separation arises \cite{gelman}. They have been designed to produce stable and regularised estimates. These priors are based on the Student's \emph{t}-distribution.

ABN is essentially a model selection technique. Indeed a large number of simple models have to be evaluated so that a comprehensive global structure can be selected. It is known that when a weakly informative prior is used, Bayesian model selection will asymptotically always prefer the simpler model, regardless of the data. This is called Lindley's paradox \cite{Lindley1957}. Using a weakly informative prior for the parameters leads to reasonable parameters estimates, compared to a pure maximum likelihood estimation for a given network.  But the main objective of ABN analysis is performing structural inference which is precisely negatively affected by weakly informative priors.

\section{Implementation and Simulation study} 
\label{sec:4}
In a practical perspective computational speed is the major concern in an ABN context as the number of models to evaluate grows super exponentially with the number of random variables. The estimation of Bayesian regression coefficients using Gibbs or Metropolis algorithms is usually not fast enough. Especially because the model selection approach is based on a point estimate of the posterior, instead of using the full network information. So an appealingly fast and reliable procedure to fit the model is described in \cite{gelman}. The procedure is an alteration of the classical iterative reweighted least squares algorithm that uses an approximate expectation-maximization algorithm to update the regression coefficients at each step. 
The prior information is taken into account through augmented data. 
This procedure is used to estimate the posterior for every possible combination of all the variables. 
The output of this procedure is a comprehensive list of scores. 
Further details for this first step are given in \cite{Krat:Furr}.
In a second step, an exact search is performed to select the network with the highest possible global score \cite{koivisto}. The simulation study has been carried out using the package \emph{abn} \cite{Kratzer} in the R software environment \cite{R}.

\subsection{Data separation}
In order to illustrate the influence of the prior on an ABN analysis, we randomly simulate BNs consisting of 10 discrete variables with 80$\%$ of the possible edges expressed. Each edge represents the same coefficient set to 0.99 on the logit scale, i.e., =\,expit(5). 
For sample sizes $N=100, 500, 1000$ and $10'000$ we randomly generate 50 distribution of the selected network. 
The two priors used are a weakly informative prior (WI) which is normally distributed with mean zero and variance 1000 and a Student's \emph{t}-prior (ST) with one degree of freedom (i.e., Cauchy) and scale parameter 2.5. Then the true positive rate (TPR) and the false positive rate (FPR) are used to measure the accuracy of the selected networks. Every selected network is transformed to an essential graph, as two networks of the same Markov class of equivalence could differ substantially in term of edges but having the same score as they represents the same assertions of conditional independence.

\begin{figure}
\centering
\includegraphics[width=\textwidth]{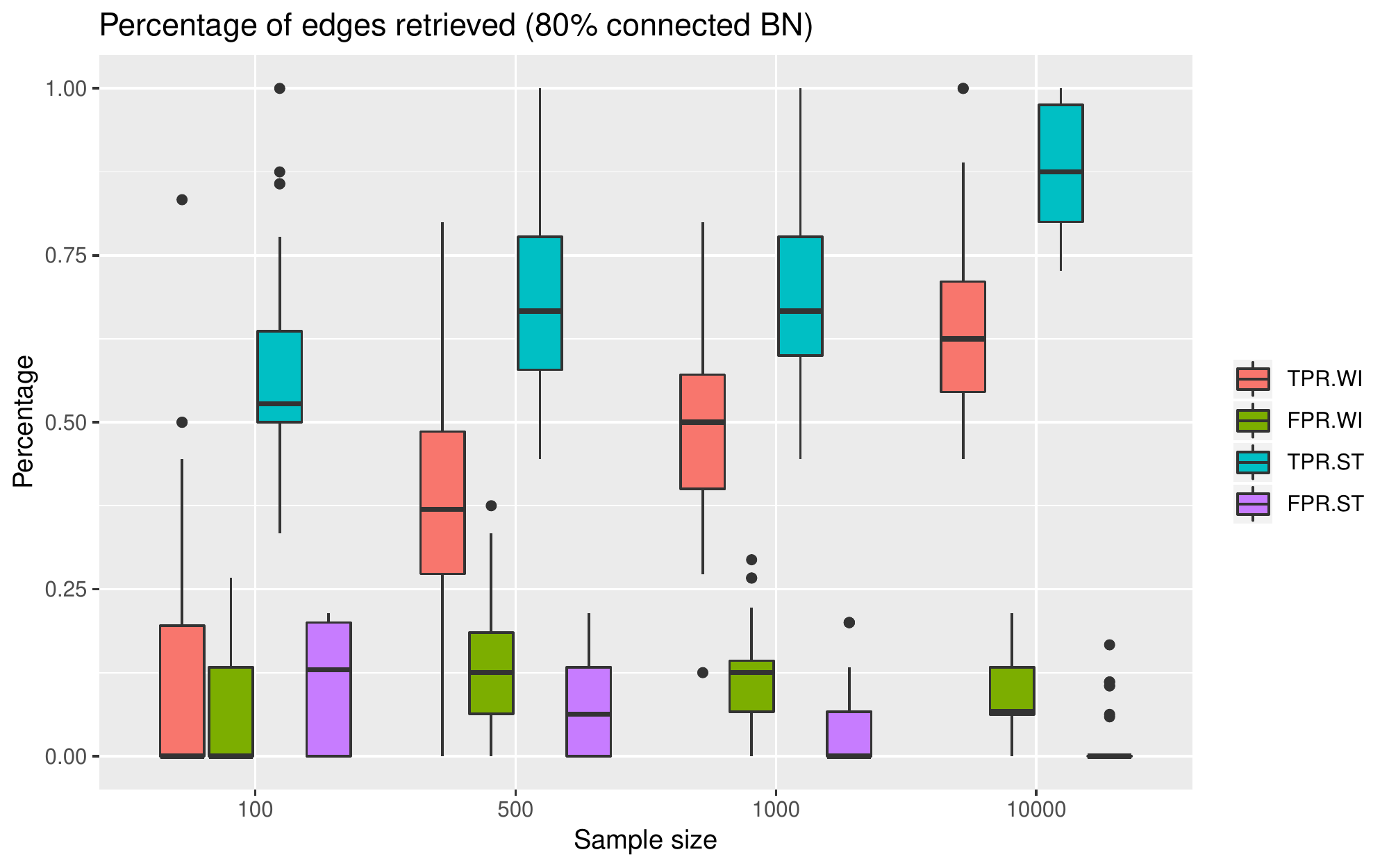}
\caption{Accuracy measures for retrieved edges for 80\% connected five nodes ($n=5$) simulated Bayesian networks as a function of data sample size ($N=100, 500, 1000$ and $10'000$).
The boxplots (each based on 50 simulations) show the true positive rate (TPR) and the false positive rate (FPR) outcome of the weakly informative prior (WI) and Student's \emph{t}-prior (ST). 
}
\label{plot} 
\end{figure}


Figure~\ref{plot} shows the TPR and FPR as a function of the sample size for two different priors and  illustrates that both priors exhibit a proper ``asymptotic`` behaviour when sample size increases:  TPR and TNR tend to 100$\%$ and 0$\%$, respectively. The chosen coefficients (0.99) of the edges in each BN leads almost surely to data separation for most of the possible variables' combination. Not surprisingly,  the Student's \emph{t}-prior has a better accuracy for network scoring for both positive and negative edges selection.

\subsection{Lindley's paradox}
An ABN modelling is based on multiple approximations such as score as proxy for selecting the best network and the scoring procedure itself. So even if Lindley's paradox is a known theoretical concern it could potentially have a limited impact in practice. 

In order to illustrate Lindley's paradox in a plausible situation, we randomly simulate BNs of $n=10$ nodes with a range of different edge densities. Each edge has a known coefficient. Then, we simulate 50 synthetic datasets of 1000 observations per network density. For this simulation study three priors have been used: the two priors described above and a strongly informative prior (SI), which is normally distributed with mean set to the true coefficient of the index edge and variance 0.1. Of course, this last prior is not realistic in practice but it is added here to illustrate the ``asymptotic'' behaviour. The average normalised number of parent is used to illustrate Lindley's paradox. For this illustration, we divide the average number of a simulated network by the true number of parents of the original network. Then, BNs are fitted using binomial regression using different priors and the essential graphs are extracted.

\begin{figure}
\begin{subfigure}{.49\textwidth}
  \hspace*{-1mm}\includegraphics[height=6.2cm]{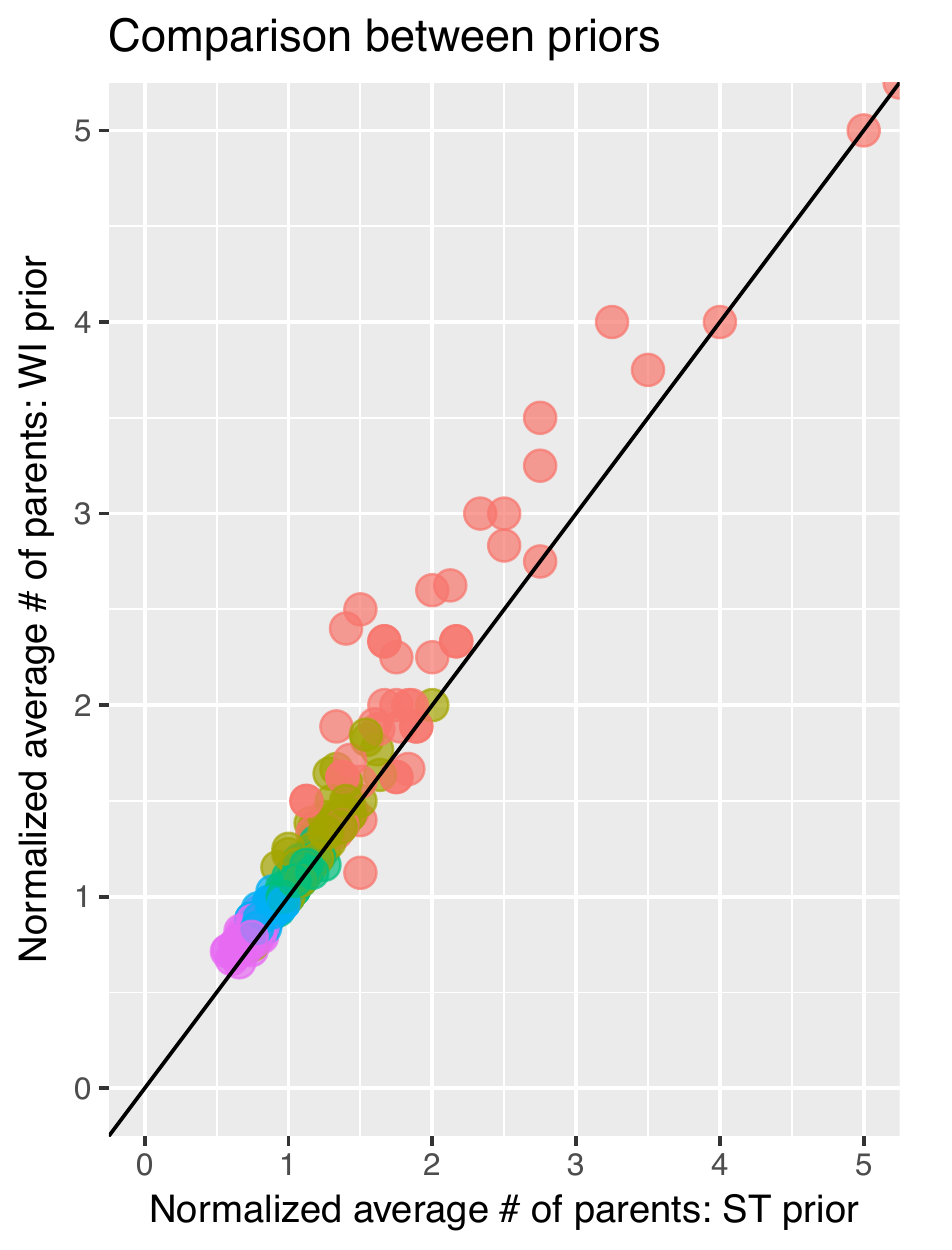}
  \caption{Student's \emph{t}-prior (ST) and the\\ weakly informative prior (WI).}
  \label{fig:sfig1}
\end{subfigure}%
\begin{subfigure}{.49\textwidth}
  \hspace*{-1mm}\includegraphics[height=6.2cm]{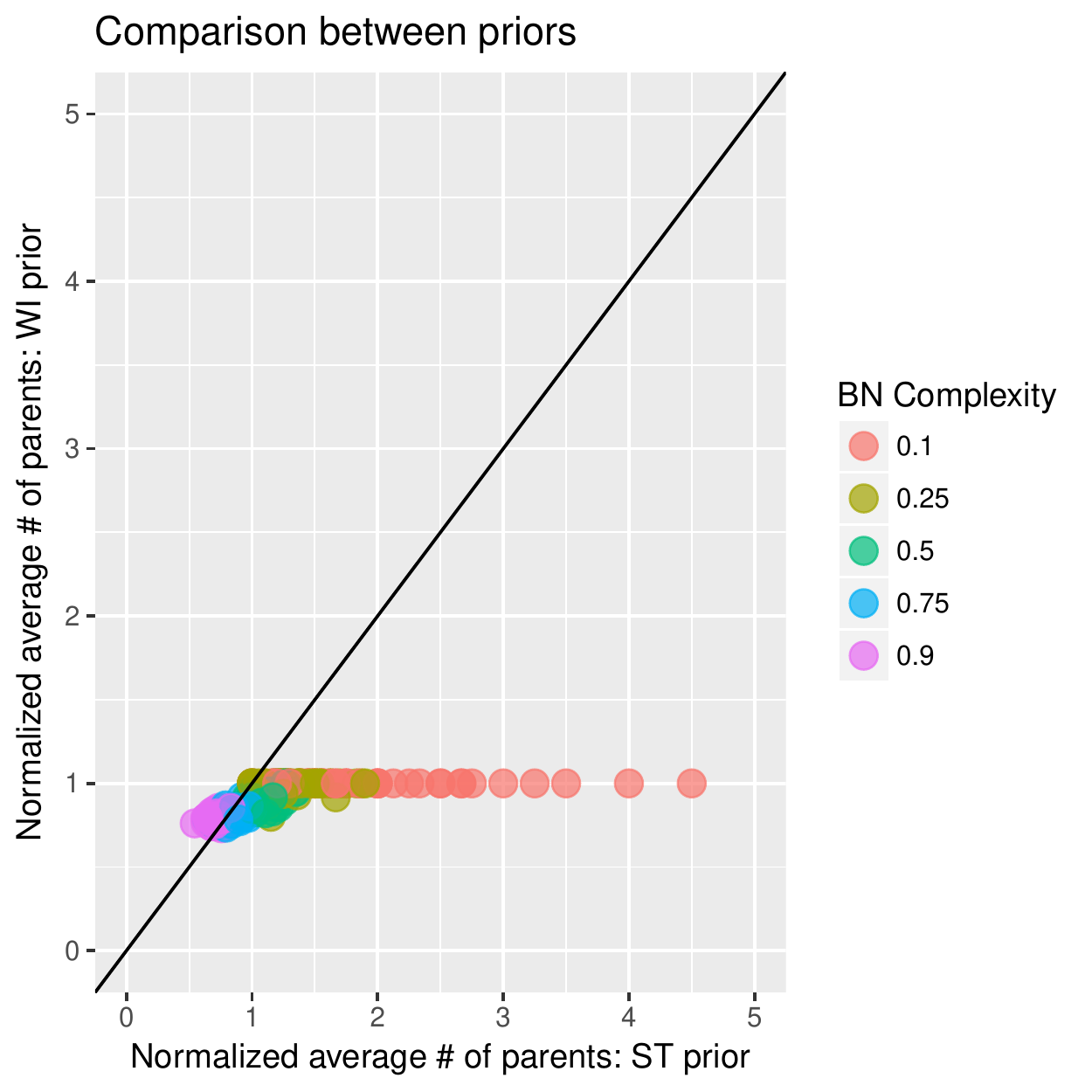}
  \caption{Student's \emph{t}-prior (ST) and the\\ strongly informative prior (SI).}
  \label{fig:sfig2}
\end{subfigure}
\caption{Comparison of different priors for different networks complexity (edge densities varying between 0.1 and 0.9).}
\label{fig}
\end{figure}

Figure~\ref{fig} summarizes the simulation result and compares the (normalized) average number of parents of the fitted BN under different priors.
If the selected DAGs are subjected to differential issue under complexity selection due to the weakness of prior information one should see a scatter plot deviating from the diagonal.   As seen in Figure~\ref{fig}(a) and (b), sparse networks, i.e. low network complexity, are more impacted by than highly connected ones. The marginal posterior likelihood seems to overfit the sparse network structure and to underfit dense networks. In Figure~\ref{fig}(a) one can see that a weakly informative prior performs comparably as Student's \emph{t}-prior. Whereas in Figure~\ref{fig}(b) the effect of highly informative prior is clearly visible. The selected networks almost never exceed in term of complexity the true networks. Surprisingly, even such a prior does not allow the scoring procedure to optimally select dense networks.

\section{Future Developments: Conjugate Priors for ABN}
\label{sec:5}

In Section \ref{sec:4} we showed that priors play a major role in ABN modeling by (i) comparing the effect of different priors on data separation when dealing with discrete data and (ii) differential density network selection depending on the prior information. 
The simulation study highlights the need to further study suitable priors for ABN modeling. In settings with sparse data or with data separation, we believe that  a conjugate prior is appealing and addressing several issues. 
If the conjugacy property will be satisfied in the context of ABN, it would lead to evident benefits. For example, a closed form distribution for the posterior might be available.
This result would lead to huge advantages in terms of the marginal likelihood computation by reducing the time for the model selection process. Similarly, the parameters estimation will also benefit from this choice.
Moreover, it should also be possible to tackle the score equivalence problem, typical of the BN literature \cite{Heckerman1995,Pittavino2016}.

In order to achieve this goal, we consider the link between ABN models and GLMs and exploit features of the exponential family.
A good candidate for this purpose is the conjugate prior distribution that belongs to a flexible family of priors called the Diaconis--Ylvisaker conjugate priors \cite{Diaconis1979}. This prior distribution was introduced by \cite{Chen2003}. A change of variables and the resulting properties applied to the Jacobian need to be checked as in \cite{Gut1995} in order to apply this distribution to our specific case.
Further work will be conducted in this direction in order to formally verify all the desirable assumptions. Additionally,  the R package \emph{abn} \cite{Kratzer} should be equipped with further priors for practical usage and  availability  for the statistical community.

\bibliographystyle{spmpsci}
\input{ABNprior_references}


\end{document}

%% file: ABNprior_references.tex
%
%
%

%% file: kratzer_furrer_pittavino.bbl
\begin{thebibliography}{99.}%
%
%



\bibitem{Jansen2003}
Jansen, R., Yu, H.Y., Greenbaum, D., Kluger, Y., Krogan, N.J., Chung, S.B.,
  Emili, A., Snyder, M., Greenblatt, J.F., Gerstein, M.: A Bayesian networks
  approach for predicting protein-protein interactions from genomic data.
\newblock Science \textbf{302}(5644), 449--453 (2003)

\bibitem{Dojer2006}
Dojer, N., Gambin, A., Mizera, A., Wilczynski, B., Tiuryn, J.: Applying dynamic
  Bayesian networks to perturbed gene expression data.
\newblock BMC Bioinformatics \textbf{7}, 249 (2006)

\bibitem{Poon2007}
Poon, A.F.Y., Lewis, F.I., Pond, S.L.K., Frost, S.D.W.: Evolutionary
  interactions between n-linked glycosylation sites in the HIV-1 envelope.
\newblock PLOS Computational Biology \textbf{3}(1), e11 (2007)

\bibitem{Djebbari2008}
Djebbari, A., Quackenbush, J.: Seeded Bayesian networks: Constructing genetic
  networks from microarray data.
\newblock BMC Systems Biology \textbf{2}, 57 (2008)


\bibitem{Hodges2010}
Hodges, A.P., Dai, D.J., Xiang, Z.S., Woolf, P., Xi, C.W., He, Y.Q.: Bayesian
  network expansion identifies new ROS and biofilm regulators.
\newblock PLOS One \textbf{5}:3, e9513 (2010)

\bibitem{Lewis2012} 
Sanchez-Vazquez, M.J., Nielen, M., Edwards, S.A., Gunn, G.J. and Lewis F.I.: 
Identifying associations between pig pathologies using a 
multidimensional machine learning methodology. BMC  Veterinary Research \textbf{8}:151, 1--11 (2012)


\bibitem{Hartnack2016} 
Hartnack, S., Springer, S., Pittavino, M. and Grimm, H.: Attitudes of Austrian veterinarians towards euthanasia in small animal practice: impacts of age and gender on views on euthanasia. BMC  Veterinary Research \textbf{12}, 1--14 (2016)

\bibitem{Pittavino2017a} 
Pittavino, M., Dreyfus, A., Heuer, C., Benschop, J., Wilson, P., Torgerson, P., Furrer, R.: Comparison between Generalized Linear Modelling and Additive Bayesian Network. Identification of Factors associated with the Incidence of Antibodies against Leptospira interrogans sv Pomona in Meat Workers in New Zealand. Acta Tropica \textbf{173}, 191--199 (2017)


\bibitem{Rijmen2008} Rijmen, F.: Bayesian networks with a logistic regression model for the conditional probabilities. International Journal of Approximate Reasoning \textbf{48}:2, 659--666 (2008)

\bibitem{zorn2005solution}
Zorn, Ch.: A solution to separation in binary response models. Political Analysis \textbf{13}:2, 157--170 (2005)

\bibitem{firth1993bias}
 Firth, D.: Bias reduction of maximum likelihood estimates. Biometrika \textbf{80}:1, 27--38 (1993)

\bibitem{gelman}
 Gelman, A., Jakulin, A., Pittau, M. G., and Su, Y. S.: A weakly informative default prior distribution for logistic and other regression models. The Annals of Applied Statistics \textbf{2}:4, 1360--1383 (2008)

\bibitem{Lindley1957}
Lindley, D.V.: A statistical paradox. Biometrika \textbf{44}:1--2, 187--192 (1957)

\bibitem{Krat:Furr}
Kratzer, G. and Furrer, R.:
    Information-theoretic scoring rules to learn additive Bayesian network applied to epidemiology.
    arXiv:1808.01126 (2018)

\bibitem{koivisto}
Koivisto, M., and Sood, K.: Exact Bayesian structure discovery in Bayesian networks. Journal of Machine Learning Research 5(May), 549--573 (2004)



\bibitem{Kratzer}
Kratzer, G., Pittavino, M., Lewis, F., Furrer, R.: abn: an R package for modelling multivariate data using additive Bayesian networks. R package version 1.2. https://CRAN.R-project.org/package=abn (2018)


\bibitem{R} 
R Development Core Team. R: a language and environment for statistical computing.
R Foundation for Statistical Computing, Vienna, Austria. http://www.R-project.org (2017)

\bibitem{Heckerman1995}
Heckerman, D., Geiger, D. and Chickering, D. M.: Learning Bayesian networks: the combination of knowledge and statistical data. Machine Learning \textbf{20}:3, 197--243 (1995).

\bibitem{Pittavino2016} 
Pittavino, M.: Additive Bayesian Networks for Multivariate Data: Parameter Learning,
Model Fitting and Applications in Veterinary Epidemiology. PhD Thesis. University of Zurich (2016)



\bibitem{Diaconis1979}
Diaconis, P. and Ylvisaker, D.: Conjugate priors for exponential families. The Annals of Statistics \textbf{7}:2, 269--281 (1979)


\bibitem{Chen2003}
Chen, M. and Ibrahim, J.: Conjugate priors for generalized linear models. Statistica Sinica \textbf{13}:389, 391, 461--476 (2003)

\bibitem{Gut1995}
Guti\'errez-Pe\~na, E., and Smith, A. F. M.: Conjugate parameterizations for natural exponential families. Journal of the American Statistical Association \textbf{432}:90, 1347--1356 (1995)






%
%

%

%
\end{thebibliography}
